\documentclass[12pt]{article}
\usepackage{geometry}                
\usepackage{graphicx}
\usepackage{amssymb}
\usepackage{epstopdf}
\usepackage{bigdelim}
\usepackage{amsmath}
\usepackage{wrapfig}
\usepackage{hyperref}
\usepackage{color}

\begin{document}

 \title{Tensile Strength and the Mining of Black Holes} 

\author{Adam~R.~Brown   \vspace{.1 in}\\
\textit{\small{Princeton Center for Theoretical Science, Princeton, NJ 08544, USA}} }

\date{}

\maketitle

\begin{abstract}

There are a number of important thought experiments that involve raising and lowering boxes full of radiation in the vicinity of black hole horizons. This paper looks at the limitations placed on these thought experiments by the null energy condition, which imposes a fundamental bound on the tensile-strength-to-weight ratio of the materials involved, makes it impossible to build a box near the horizon that is wider than a single wavelength of the  Hawking quanta and puts a severe constraint on the operation of `space elevators' near black holes. In particular, it is shown that proposals for mining black holes by lowering boxes near the horizon, collecting some Hawking radiation and dragging it out to infinity cannot proceed nearly as rapidly as has previously been claimed and that as a consequence of this limitation the boxes and all the moving parts are superfluous and black holes can be destroyed equally rapidly by threading the horizon with strings.

\end{abstract}


\section{Introduction}

Classical black holes live forever. The area theorem shows that not only can black holes not be destroyed, their horizon area cannot decrease at all \cite{Hawking:1971tu}. Though Penrose-style processes can extract energy stored in electric, magnetic, or gravito-kinetic fields outside the horizon of charged or spinning black holes \cite{Penrose:1969pc,Christodoulou:1972kt}, no energy can be extracted from the hole itself, and once the charge/spin is gone we are left with a Schwarzschild black hole that it both inert and eternal.

Quantum black holes, however, disintegrate into Hawking radiation. But black hole evaporation is slow. A 3+1-dimensional Schwarzschild black hole of mass $M$ self-destructs in a time \cite{Hawking:1974rv}
\begin{equation}
\textrm{lifetime} = \frac{ 5120 \pi \, b\,  G^2  }{\hbar \, c^4} M^3,
\end{equation}
which for a solar mass black hole  comes to $10^{57}$ times the age of the Universe\footnote{This formula is correct for large semiclassical black holes $M\gg (8 \pi G)^{-1/2} \equiv M_{Pl}$, which is the only limit in which the notion of a black hole is well defined, and the only limit we will be interested in in this paper. The O(1) constant $b$ depends on the number and nature of the massless species, and on the greybody corrections to the Stefan-Boltzmann law \cite{Page:1976df}. In the rest of this paper we will use Planck units so that $G = \hbar = c = 1$.}. Can the hole be made to relinquish its energy sooner? 

Unruh and Wald have argued that it can \cite{Unruh:1982ic}. They have argued that by lowering a box down close to the horizon, filling it with Hawking radiation and raising the box back out to infinity, that the black hole can be stripped of its thermal atmosphere and destroyed in a time that scales like the Schwarzschild time $M$ \cite{UnruhMiningFast:1983}. I will show that  implemented literally the standard prescription will result in the black hole horizon swelling and consuming the box, so that rather than using a box to rob the black hole of its radiation, the black hole instead robs us of our box. Though black holes can still be destroyed parametrically faster than by evaporation, the constraints (Eqs.~\ref{lifetime:mu} \& \ref{lifetime:radiation}) are parametrically slower than those previously derived. Indeed, I will argue that the limitations on the mining of black holes by boxes are so severe that no advantage (and some disadvantage) is gained by using boxes, and we are better off mining the black hole by just threading the horizon with strings in the manner of Lawrence \& Martinec \cite{Lawrence:1993sg} and of Frolov \& Fursaev \cite{Frolov:2000kx,Frolov:2002qd}.

The strongest constraints are going to come from limitations the energy conditions (null, weak, and dominant all agree in this case) place on the properties of the apparatus we are going to use to do the mining. The mining of black holes is meant to be a quasistatic process, so we can use  versions of these conditions averaged over semiclassical distances. The averaged null energy condition is the most permissive, and therefore the most secure, of all energy conditions. It is obeyed by all observed matter, classical or quantum, and it seems doubtful that there is any notion of black hole thermodynamics that could be salvaged if we allow our equipment to violate it. The energy conditions demand that the tension $T$ of a static rope cannot exceed its mass-per-unit-length $\mu$
\begin{equation}
\mu \geq T. \label{eq:nullenergycondition}
\end{equation}
A rope that is tense must also be dense\footnote{This fundamental limit, $T/\mu= c^2 = 9 \times10^{16}N/(kg/m)$ in SI units, far exceeds the breaking point of any material that derives its strength from interatomic forces (e.g. defect-free carbon nanotubes can sustain no more than $T/\mu=10^{-8}$). `Ropes' that saturate the condition have no longitudinal rest frame; examples include electric field lines, flux tubes, and cosmic and fundamental strings.}. Subject to a greater force, the rope must stretch or the rope must break; what the rope cannot do is resist.  Let's examine the consequences.

\section{Tensile Strength} \label{sec:TS}

 A general static spherically-symmetric spacetime has metric
\begin{equation}
ds^2 = - \chi(r)^2 dt^2 + \frac{dr^2 }{f(r)^2} + r^2 d\Omega_2^2 . \label{eq:generalmetric}
\end{equation}
For a Schwarzschild black hole, $\chi = f= (1 - 2M/r)^{1/2}$. 
 A general static spherically-symmetric matter distribution has stress-energy
\begin{equation}
T^{\mu}_{\ \, \nu} = \textrm{diag} \{- \rho, p_r ,p_\theta, p_\theta \} = \frac{1}{4 \pi r^2} \textrm{diag} \{- \mu, -T ,S,S \} , \label{eq:generalmatter}
\end{equation}
where $\mu(r)$ is the mass-per-unit-radial-length, $T(r)$ is the radial tension and $S(r)$ is the angular compression stress.  The condition for equilibrium is $\nabla_{\mu} T^{\mu}_{\ \, r} = 0$, or 
\begin{equation}
\frac{dT}{dr} + \frac{1}{\chi} \frac{d \chi}{dr}   T + \frac{2}{r} S = \frac{1}{\chi} \frac{d \chi}{dr}  \mu . \label{eq:underpressure}
\end{equation}
The weight of the material (the RHS) must be supported by increasing the radial tension (the first term), holding the radial tension fixed while the tension below redshifts away (the second term), or by angular compression stress (the third term). 

The null energy condition (NEC) requires  that $T^{\mu}_{\ \nu} k_{\mu}k^{\nu} \geq 0$ for every null vector $k^{\mu}$. Choosing the radial null vector reproduces $\mu > T$. Choosing the angular null vector requires that $\mu >- S$. Let's investigate the implications of this condition as it pertains to ropes and boxes near black holes.

\paragraph{1. Ropes must be heavy.} A hanging rope has radial tension ($T>0$) but no angular stress ($S=0$). If a rope of constant $\mu$ is suspended  from infinity down towards the black hole horizon, what tension in the rope is required to keep it static? Eq.~\ref{eq:underpressure} tells us that the required tension is independent of $r$ and independent of $M$ 
\begin{equation}
T(\mu,r, M) = \mu.
\end{equation}
Closer to the horizon the gravitational field $g \equiv d  \log \chi / ds$ is stronger, but there is less rope to support below\footnote{This solution acts as a counter-example to the claim of \cite{Gibbons:1972} that there can be no such solution. (The fault enters in the paper's Eq.~5, which disregards the changing radial component of $t^a$, see 
%
%
%
%
\cite{Fouxon:2008pz} for a discussion.) It is, however, the case that there is no way \emph{other} than tensile ropes to hold something fixed near a black hole; this is the subject of Appendix A.}. The ability of a constant tension rope to support itself is a uniquely relativistic effect; in Newtonian mechanics the tension must always increase with height to compensate for the increased weight, but in  curved spacetime the weight of the rope redshifts away.

Thus for a constant-$\mu$ rope to support itself, it must saturate the NEC bound. And even a constant-$\mu$ rope that saturates the NEC bound must expend all its tensile strength supporting its own weight, if it stretches all the way down to the horizon, leaving none over to support a box. By getting rid of the rope below a certain height, we can free up tensile strength, but only enough to support a box no heavier than the weight of the excised rope. For a thin box of proper mass $\mathfrak{m} \ll M$ at a radius $r=R$ in Schwarzschild spacetime, integrating Eq.~\ref{eq:underpressure} gives the required tension and therefore the required density in the rope as $\mathfrak{m} g|_{r=R}$, or
\begin{equation}
\mu \geq T =  \frac{1}{\chi} \frac{M \mathfrak{m} }{r^2 } \biggl|_{r=R}  \label{eq:Tfixedgravity}.
\end{equation}

It is sometimes said that the force required at infinity to hold a box fixed near a black hole remains bounded even for masses arbitrarily close to the horizon: though the gravitational field gets ever stronger, the redshifted gravitational force remains finite. This is technically true, but misleading. 
The NEC demands that the rope be heavy (Eq.~\ref{eq:Tfixedgravity}), which means that by the time you are far from the hole very little of the tension is devoted to supporting the weight of the box, and almost all is devoted to supporting the weight of the rope. The force at infinity required to suspend both box \emph{and} supporting NEC-obedient rope does diverge as the horizon is approached.

\paragraph{2. Boxes must be heavy.} A second consequence of the NEC is that a box full of massless radiation must weigh parametrically as much as its contents just to stop them escaping into the vacuum. We can see this already in the flat spacetime $\chi = f = 1$ limit. It is the angular tension $S<0$ in the material of a balloon that resists the pressure of its contents. Massless radiation of density $\rho$ has pressure $p=\frac{1}{3} \rho$, so the inside edge of the balloon surface is under tension $T = 4 \pi R^2 \frac{1}{3} \rho$, while  the outside edge is exposed to the vacuum $T=0$. We can then integrate Eq.~\ref{eq:underpressure} across the thin surface at $r=R$ to derive
\begin{equation}
4 \pi R^2 \frac{1}{3} \rho = - \Delta T = - \frac{2}{R} S \Delta r \leq \frac{2}{R} \mu \Delta r = \frac{2}{R} M_{\textrm{box}}.
\end{equation}
The mass of the radiation is $\frac{4 \pi}{3} \rho R^3$ so that\footnote{If the limit of the stability against radial perturbations is $w_{\theta} \equiv S/\mu = - 1/2$, as the results of \cite{Brady:1991np} might lead one to speculate, then there is a tighter bound $M_{\textrm{box}} \geq M_{\textrm{radiation}}$. See also \cite{Bousso:2002bh}.}, independent of the number of species, 
\begin{equation}
M_{\textrm{box}} \geq \frac{1}{2} M_{\textrm{radiation}} . 
\end{equation}

\paragraph{3. Boxes must be narrow.}
A third consequence of the NEC is that a single box hanging from a single rope near a black hole horizon can be no wider than the local Hawking wavelength. To see this, let us first see what constraints the NEC places on ropes suspended not from one point, as in the previous discussion, but from two. 

In Newtonian mechanics, the profile adopted by a constant density rope suspended from two points in a constant gravitational field is the catenary $y=  y_b-a+a  \cosh (x-x_b)/a$, where $a$, $x_b$ and $y_b$ are determined by the boundary conditions. The tension at the bottom is $\mu g a$. This Newtonian intuition suggests that the  NEC will bound the radius of curvature at the bottom, $a$, to be less than about $g^{-1}$. 

Let us consider the region just outside a fixed Schwarzschild horizon. Using $f = \chi = (1 - 2M/ r)^{1/2}$ and moving to the near horizon limit ($\chi \ll 1$) of Eq.~\ref{eq:generalmetric} gives Rindler coordinates
\begin{equation}
ds^2 = - \chi^2 dt^2 + 16M^2 d \chi^2 + 4M^2 \left( d \theta^2 + \cos^2 \theta d \phi^2 \right). \label{eq:nearhorizon}
\end{equation}
The horizon lies at $\chi =0$, where the  local gravitational field $g = 1 / (4\chi M)$ diverges. 
The action of a static constant-$\phi$ constant-$\mu$ NEC-saturating string hanging with shape $\chi(\theta)$ is proportional to 
\begin{equation}
S  \sim  \mu  \int d t \, d \theta \, \cos \theta  \, \chi \sqrt{1 + 4 \chi'(\theta) ^2 }.
\end{equation}
If we consider strings that are not only close to the horizon but also take up a small angular scale we can treat the $\cos \theta$ as essentially fixed so that we are treating the horizon as a plane. Then the action provides the same function to be extremized as the potential energy in the Newtonian case, and the solution is
\begin{equation}
\chi(\theta) = \frac{1}{2} \chi_0 \cosh \frac{\theta}{\chi_0}.  \label{eq:catenaryform}
\end{equation}
Remarkably, the shape of a constant-$\mu$, constant-$T$ string in the non-constant gravitational field of a black hole has the same functional form as the shape of a constant-$\mu$, non-constant-$T$ rope in a constant Newtonian potential \cite{Friess:2006rk,Berenstein:2007tj,deKlerk:2011ua}. The difference is that in this case we have specified the tension so if we fix the point of closest approach to the horizon, we also fix the radius of curvature there. Since we can make $T$ less than $\mu$, we can make the rope hang steeper than Eq.~\ref{eq:catenaryform}. Since we cannot make $T$ greater than $\mu$, we cannot make the rope hang shallower than Eq.~\ref{eq:catenaryform}. (Appendix D confirms this intuition.)


The implication of this is that two points at the same $\chi$ can be connected by a NEC-satisfying string only if they are sufficiently close. If they are separated by a distance $2M \Delta \theta$, then a string can be hung between them only if there is some solution to the above equation. Minimizing $\chi(\theta)$ with respect to $\chi_0$ gives the bound
\begin{equation}
\Delta \theta < (1.3254 \ldots) \chi . \label{eq:separationlimitstring}
\end{equation}

We can repeat this analysis for the rotationally symmetric box formed by hanging a NEC-saturating sheet from a circular support. In the near-horizon, small-angular-size, fixed-Schwarzschild limit the action is proportional to 
\begin{equation}
S \sim \int d \theta \ \chi \ \theta \sqrt{1 + 4 \chi'(\theta)^2 } ,
\end{equation}
so that the shape must satisfy
\begin{equation}
\frac{2\chi''}{1 + 4 \chi'^2} =  \frac{1}{2 \chi} - \frac{2 \chi'}{\theta} .\label{equationofmotion}
\end{equation}
The dilation symmetry ensures that if $\chi(\theta)$ is a solution, so is $\lambda \chi(\theta/\lambda)$, just as for the case of the NEC-saturating string. 
The shape equation can be numerically integrated to reveal that a sheet can be hung from a circle without breaking only if
\begin{equation}
\Delta \theta < (2.1754 \ldots) \chi . \label{eq:separationlimitbrane}
\end{equation}
(Again, Appendix \ref{appendix:stringcaternary} confirms that this is the widest any NEC-satisfying box can be, not just the widest a NEC-saturating box is.)

At a height $4 M \chi$ above the horizon the wavelength of a Hawking photon is approximately $M \Delta \theta \sim 4 M \chi$, so the NEC permits us to build boxes only just wide enough to fit a single wavelength of Hawking radiation, no matter whether we use ropes or sheets. If we employ many supporting strings, then we can hang many boxes that collectively cover a larger area, but no single box with a single point of support can have a width that exceeds the local Hawking wavelength. For every Hawking wavelength we wish to enclose, we need another supporting rope.

\paragraph{Backreaction and melting.}
If the rope is too heavy, it will undergo gravitational collapse. Defining $f(r) \equiv (1 - 2m(r)/{r} )^{1/2}$, the $G^{t}_{\ \,t}=8 \pi T^{t}_{\ \,t}$ component of Einstein's equation for a vertical rope is that
\begin{equation}
\frac{d m(r)}{dr} = \mu.
\end{equation}
A horizon exists where $f=0$, so a uniform static string can avoid gravitational collapse only if it is lighter than half a Planck mass per Planck length,
\begin{equation}
\mu < \frac{1}{2}. \label{eq:backreactionofstring}
\end{equation}

The rope cannot be too light, though. If the rope is so slight, or the temperature so great, that the weight of even a single Hawking photon exceeds the rope's carrying capacity, Eq.~\ref{eq:Tfixedgravity}, then the rope cannot prevent the photon falling into the hole and carrying at least part of the rope with it. Since at a redshift $\chi$ a typical Hawking quantum has an energy $\mathfrak{m} \sim (\chi M)^{-1}$, the string can bear the photon only when
\begin{equation}
\mu > \frac{1}{(\chi M)^2}. \label{eq:stringmelting}
\end{equation}
For a fundamental string, this corresponds to the redshift at which the Hawking radiation has reached the melting point of the string, the Hagedorn temperature \cite{Hagedorn:1965st,Susskind:1993ki}. (This failure mode should be thought of as the string melting, not the string breaking. A NEC-saturating string breaks by Schwinger pair producing endpoints, which is a failure mode in of itself, and is the reason QCD flux tubes are of no use for our purposes---though they successfully saturate the NEC bound by having their tension equal to their energy density, a stretched flux tube pair produces pions and falls apart. The breaking problem can be evaded by finding a rope with a high endpoint mass or a low coupling. Melting, by contrast, is a problem that afflicts even unbreakable strings, and corresponds to absorbing energy from the heat bath and using it to manufacture new string so that the rope stretches uncontrollably.)

The twin constraints of backreaction and melting limit the usefulness of ropes less tense than the NEC bound. We have already seen that only a $T=\mu$ rope can support itself if it is to have constant density, so a $T<\mu$ rope must be tapered, its profile sculpted to be thicker at the top and thinner at the bottom. 
If we input $T(r) = -w \mu(r)$ into Eq.~\ref{eq:underpressure} then the solution is
\begin{equation}
\mu (\chi) = \mu_\infty \,  \chi^{-\frac{1 + w}{w}}  \ \ . 
\end{equation}
The string tapers from finite linear density $\mu_{\infty}$ at infinity to zero linear density at the horizon. 
On the one hand, the rope must not be so thick at infinity as to induce gravitational collapse, so $\mu_{\infty} < 1/2$. On the other hand, the rope must not be so thin at the black hole end that it melts and loses control of the box. A rope that is as thick at infinity as is consistent with backreaction melts at a redshift
\begin{equation}
\frac{1}{\chi^2 M^2} \sim \frac{1}{2}  \chi^{-\frac{1 + w}{w}}  \rightarrow \chi \sim M^{ \frac{2w}{1 - w}}.
\end{equation}
But most thought experiments take place down at a redshift that scales as $\chi \sim M^{-1}$ so that the temperature stays fixed even as the black hole is taken very large and semiclassical. It is only at these redshifts that the Generalized Second Law may be in jeopardy   \cite{Penrose:1969pc,Bekenstein:1972tm,Bardeen:1973gs,Bekenstein:1973ur, Bekenstein:1974ax,Bekenstein:1980jp,Bekenstein:1983iq,Bekenstein:1999bh}, only at these redshifts that buoyancy produced by the Hawking atmosphere becomes significant  \cite{Unruh:1982ic,Unruh:1983ir,Marolf:2002ay}, and only at these redshifts that we could hope to mine enough energy that the lifetime scales as the light-crossing time $M$ \cite{UnruhMiningFast:1983}.  So only a NEC-saturating string will do. Not twine, not steel, not nanotubes---to reach interesting redshifts requires our `rope' to have the maximum possible tensile strength permitted by the laws of nature. 

\section{The Destruction of Black Holes} \label{sec:TDoBH}

 Unaided Hawking radiation releases approximately one quantum  per light-crossing time $M$. If in the same time $N$ such quanta could be liberated, the lifetime would fall to
\begin{equation}
\textrm{lifetime} \sim \frac{M^3}{N}  \label{eq:species}.
\end{equation}

Near the horizon of a black hole the metric is given by Eq.~\ref{eq:nearhorizon}, the area remains fixed at $M^2$ but every other length scale is given by $\chi M$: the distance to the horizon, the wavelength of Hawking quanta, and the time spacing with which they arrive.  Every locally-measured time $\chi M$ a photon of wavelength $\chi M$ passes through each cell of area $(\chi M)^2$. Due to the gravitational time dilation near the horizon, a locally measured time of $\chi M$ corresponds to an asymptotically-measured time of $M$. Thus through a given angular area element near the horizon, the number of photons passing per asymptotically-measured light-crossing time is 
\begin{equation}
N = \frac{\textrm{area}}{\chi^2 M^2}. \label{eq:availablephotons}
\end{equation}
Taking the area to be a whole sphere surrounding the hole, this number of photons is $N \sim 1/\chi^2$. What this means is that almost all photons that make it past a sphere of redshift $\chi \ll 1$ do not make it out to infinity. This effect can be understood already in the geometric optics approximation (marginally applicable here because the wavelength and redshifting-doubling-length coincide). As a matter of geometry, null rays near the horizon must be aimed within an angle $\Delta \psi \sim \chi$ of the vertical in order to escape the hole---any greater deviation and they loop round and are recaptured by the hole. Angular momentum makes escaping the hole more difficult (c.f. Eq.~\ref{eq:angularmomentumbarrier}).

The mining proposal \cite{Unruh:1982ic,UnruhMiningFast:1983,Davies:1984qb} is that we reach in with a box and help these photons over the angular momentum barrier. No matter whether it expends its own energy climbing out of the gravitational potential, or we have to expend energy dragging it out, the net energy we recover from a photon is $\chi$ times the proper energy it had when we captured it, which is to say $\chi \frac{1}{\chi M} = \frac{1}{M}$.  It's not that we recover more energy per photon, it's rather  that we recover more photons. How many can be liberated per time $M$ is going to be determined, through Eq.~\ref{eq:availablephotons}, by how deep we can mine. We turn to that question now.  \\

If the mass-per-unit-length of an individual NEC-saturating rope is $\mu_s$, and the total number of ropes we wish to deploy is $N_s$, then we have the following constraints. 

\emph{Constraint 1: melting}. If we are to scoop from a redshift $\chi$, then the strings cannot have melted at that depth. Equation~\ref{eq:stringmelting} then implies
\begin{equation}
\frac{1}{\chi^2 M^2} < \mu_s. \label{constraint:melting}
\end{equation}

\emph{Constraint 2: gravitational backreaction}. The collective mass of the quasistatic strings must not induce gravitational collapse. Equation~\ref{eq:backreactionofstring} then implies
\begin{equation}
\mu_s N_s < 1/2.  \label{constraint:backreaction}
\end{equation} 

\emph{Constraint 3: box width.} 
Equations~\ref{eq:separationlimitstring} and \ref{eq:separationlimitbrane} show that it is impossible to construct a box that is wider than the local wavelength of the radiation, so it is impossible for a single rope to lift parametrically more than a single Hawking quantum per light-crossing time
\begin{equation}
N < N_s. \label{constraint:oneeach}
\end{equation}

One lower bound on the lifetime comes from combining  Eqs.~\ref{constraint:backreaction}~\&~\ref{constraint:oneeach}. If we have at our disposal a NEC-saturating rope with a given fixed $\mu_s$, then gravitational backreaction limits the number of ropes we may deploy, so the black hole can be destroyed in a time no shorter than
\begin{equation}
\boxed{\textrm{lifetime} \geq \mu_s M^3}. \label{lifetime:mu}
\end{equation}
This is a factor of $G \mu_s$ shorter than the unaided evaporation time. If we have a number of different weights of suitable NEC-saturating rope at our disposal, this lower bound indicates we should choose the one with the smallest $\mu_s$. But only up to a point. If the string is too light then it melts before it can get deep enough. If we wish to collect many photons then we must reach deep, but deep means hot. Equation \ref{constraint:melting} implies that if we have complete freedom to pick $\mu_s$, then the optimal tradeoff between backreaction and melting is given by picking $\mu_s \sim M^{-1}$. This gives a second lower bound  on the lifetime\footnote{We have derived these lifetime lower bounds by considering the backreaction and melting of the supporting strings. We could equally well have considered the backreaction of the box. If the box is constructed out of a lattice of strings, then the lattice spacing cannot exceed the wavelength of the photons it is to contain $\chi M$, so the proper mass of the box must be at least 
\begin{equation}
\mathfrak{m} \geq \frac{\mu \, \textrm{area}}{\chi M} . \label{boxofstringsweight}
\end{equation}
Appendix \ref{Appendix:backreaction} constructs the metric for the complete quasistatic mining configuration, including full backreaction, and gives in Equation~\ref{eq:boxcollapse}  the condition under which a quasistatic box suspended near a black hole will undergo gravitational collapse; this recovers Eq.~\ref{lifetime:mu} and, coupled with Eq.~\ref{constraint:melting}, recovers Eq.~\ref{lifetime:radiation}. (Using boxes constructed of branes instead of strings is counterproductive. The mass of such a box is $\mathfrak{m} \sim \textrm{area} \times \textrm{brane tension}$, which is heavier than Eq.~\ref{boxofstringsweight} at large $M$, and therefore is more vulnerable to gravitational collapse.) Yet a third way to derive the constraint of Eq.~\ref{lifetime:radiation} is to consider the backreaction of the radiation being mined. The mined radiation has $\mu = \textrm{area} \times \rho =  M^2 (\chi M)^{-4} = \chi^{-4} M^{-2}$ \cite{Page:1982fm}. Requiring that this be less than $1/2$ gives $\chi > {1}/{\sqrt{M}}$ which through $N \sim 1/\chi^2$ gives Eq.~\ref{lifetime:radiation}. Similar arguments will appear in \cite{AMPS:soon}.}
\begin{equation}
\boxed{\textrm{lifetime }\geq M^2}. \label{lifetime:radiation}
\end{equation}
For intermediate mass black holes, this may be more restrictive than Eq.~\ref{lifetime:mu}. \\

The lower bounds Eqs.~\ref{lifetime:mu} and \ref{lifetime:radiation} limit the rate at which black holes can be mined with boxes. But boxes are not the only ways to mine black holes   \cite{Frolov:2000kx,Frolov:2002qd}. Lawrence and Martinec {\cite{Lawrence:1993sg} showed that, even without a box, a string dangled into a black hole wicks away Hawking radiation. The equation for perturbations on the string is just that of the s-wave bulk Hawking mode, and since the s-wave bulk mode carries away the majority of the energy in conventional Hawking radiation \cite{Page:1976df}, so a single string carries away parametrically as much energy as the whole bulk Hawking emission (`black holes radiate mainly on the brane' \cite{Emparan:2000rs,deBoer:2008gu}). Photons with high angular momentum cling to the string, deposit their angular momentum and are channeled up to infinity. Frolov and Fursaev \cite{Frolov:2000kx} argued that employing many strings black holes can be systematically mined.

Black hole mining with strings is slow: each string can carry away just one quantum per light-crossing time. But the foregoing analysis shows that the narrow boxes demanded by the NEC can do no better. The constraints on the number of strings that can support boxes carry directly over to constraints on the number of strings that can be dangled into the horizon, so the rates of these two types of mining are parametrically identical\footnote{Frolov and Fursaev \cite{Frolov:2000kx} considered multiple strings sticking into black holes and achieved the same limits on mining as is captured in Eqs.~\ref{lifetime:mu} and \ref{lifetime:radiation}. They derived the limit Eq.~\ref{lifetime:radiation} by considerations of string reconnection: the strings are safe from reconnecting with one another and being expelled from the hole if kept more than one string length apart. However, it seems like there are other ways the strings could be safe from reconnection: for example, they could be oriented strings. Happily, we have seen other ways to derive the same limit. I thank Don Marolf for discussions on this point.}.  

When it comes to black hole mining, then, boxes are superfluous. Worse than superfluous, they are an encumbrance: the many extraneous moving parts present more failure modes and burden the mining process with unnecessary $O(1)$ overheads. Rather than scoop the Hawking radiation up with boxes, both the simplest and the fastest way to destroy black holes is to puncture the horizon with a large number of NEC-saturating strings, allowing Hawking modes even of high angular momentum to flow up and away along a soaring multitude of celestial brane drains. \\

There is a great trove of energy stored in the thermal atmosphere of a black hole, by some measures all the energy the hole possesses. But we see this energy only faintly, in the rare Hawking quanta that make it out, and we grasp for it at our peril. To reach close to the horizon demands that our equipment be strong, the threat of gravitational backreaction demands that our equipment be light, but the null energy condition demands that that which is strong must also be heavy. 

These constraints partially sequester the atmosphere from our reach, and ensure that while mining can reduce the lifetime, for the largest black holes it can do so only in the prefactor---the lifetime in light-crossing times goes from the horizon area in Planck units to the horizon area in string units. Intriguingly, as shown in Appendix~\ref{appendix:n+1}, this is no longer true in higher dimensions. Mining in higher dimensions is much more effective, even considering the limitations discussed in this paper, and leads to a lifetime powers of $M$ shorter than evaporation---in high enough dimensions the lifetime scales slower than $M$. In high dimensions mining allows for the rapid destruction of black holes, the rapid extraction of their energy, and, so the modern view must go, the rapid recovery of their information.

\section*{Acknowledgements}
A warm thank you to Nima Arkani-Hamed, Raphael Bousso, Alex Dahlen, Steven Gubser, Igor Klebanov, Juan Maldacena,  Don Marolf, Don Page,  Douglas Stanford, Bill Unruh, Herman Verlinde, and Robert Wald.

\appendix

\section{Satellites, rockets, and Dyson spheres}

In the main text, we discussed the limits imposed by the energy conditions on ropes near black holes. But there are other ways you might imagine holding a box fixed near a black hole for the purpose of mining. In this appendix, I will examine three---orbital angular momentum, rockets, and compression structures---and show that they do not work, and that they all cease working at the same radius $r=3M$.

Orbital angular momentum does not work. The proper radial acceleration $\alpha \equiv | \dot{x}^{\mu} \nabla_\mu \dot{x}^{\nu}|$ required to maintain constant $r$ outside a Schwarzschild black hole is 
\begin{equation}
\alpha = \frac{1}{\chi} \frac{M}{ r^2}  - \chi \frac{L^2}{r^3}  + \frac{1}{\chi} \frac{M L^2}{r^4}, \label{eq:angularmomentumbarrier}
\end{equation}
which adds to the competition between the Newtonian gravitational attraction (augmented by $\chi^{-1}$) and the centrifugal repulsion (diminished by $\chi$) a relativistic attractive term between the hole and the orbital energy. Unlike in Newtonian mechanics, in general relativity the gravitational attraction famously overwhelms the centrifugal force at short distances, dooming any causal geodesic that approaches closer than the innermost circular orbit at $r=3M$ to inevitably hit the singularity \cite{Penrose:1964wq}. (That angular momentum is attractive inside $r=3M$ is precisely what made it so difficult for the Hawking quanta of high angular momentum in Sec.~\ref{sec:TDoBH} to escape, and is precisely the problem that mining is meant to alleviate.)

Rockets doubly do not work. First they cannot be quasistatic: near a black hole their fuel exponentially depletes in a time of order $M$. Second when inside the photon orbit at $r=3M$ their exhaust inevitably feeds the hole and pollutes the delicate energy accounting required for mining.

Compression structures do no work. If we build a `Dyson sphere' around a black hole, the compression strength in the struts required to buttress the sphere against gravity is derived in Eq.~\ref{eq:requiredw} to be
\begin{equation}
w_{\theta} \equiv S/\mu = \frac{M}{2\chi^2 r}. \label{eq:wrequired}
\end{equation} 
Even before the dominant energy condition is violated at $w_\theta >1$ (the weak and null energy conditions, by contrast, limit only tensile not compressive strength), a mechanical instability from superluminal perturbations emerges at $w_\theta >1/2$ and the shell disintegrates \cite{Brady:1991np}. This corresponds to $r=3M$.

The only way to hold a box close to the horizon is a rope under tension.

\section{Constraints on mining in $n+1$-dimensions} \label{appendix:n+1}

In an $n+1$-dimensional spacetime, the Schwarzschild radius $r_S$ and the black hole mass $M$ are no longer linearly related. Instead we have
\begin{equation}
\chi^2 = 1 - \frac{G M}{r^{n-2}}  \rightarrow r_S \sim M^{\frac{1}{n-2}} ,
\end{equation}
where $G \sim M_{\textrm{Pl}}^{1-n}$. Since approximately one quantum of energy $r_S^{-1}$ is released each light-crossing time $r_S$, the evaporation time is 
\begin{equation}
 \textrm{lifetime} \sim {M r_S^2} \sim {r_S^n}.
\end{equation}
The lifetime, measured in units of the Schwarzschild time, is the entropy. \\

The backreaction  constraint is that for all $r>r_S$,
\begin{equation}
m(r) =  M + \mu (r-r_S) < r^{n-2}.
\end{equation}
For $n>3$ this constraint is strongest immediately outside the horizon, where for a rope of mass-per-unit-length $\mu_s$ it gives a constraint on the number of ropes $N_s$ of
\begin{equation}
\mu = N_s \mu_s < M/r_S.
\end{equation}
Equation~\ref{lifetime:mu} becomes
\begin{equation}
\boxed{\textrm{lifetime} > \mu_s \,  r_S^3}. 
\end{equation}
In $3+1$-dimensions the lifetimes of the largest black holes scale as the same power of $r_S$ whether or not the black holes are being mined: only the prefactor is improved.  But in higher dimensions mining is much more effective---the new lifetime is smaller than the evaporation lifetime by full powers of $r_S/t_{\textrm{Pl}}$. Indeed for $n>5$ the mined lifetime grows slower than $M$.

We can also derive the higher dimensional version of Eq.~\ref{lifetime:radiation}, the lower bound operative when we have at our disposal NEC-saturating strings of arbitrarily small $\mu$. The number of photons passing through a sphere of a given $\chi$ is $N \sim 1/\chi^{n-1}$. The condition that the string not melt is still $\mu_s > 1/ \chi^2 r_S^2$.
The optimal string tension is therefore 
$\mu_s = r_S^{-4/({1 + n})}$ so 
Eq.~\ref{lifetime:radiation} becomes
\begin{equation}
\boxed{\textrm{lifetime} > r_S^\frac{3n -1}{n+1} }.
\end{equation}
These bounds apply equally to mining with boxes or mining with dipped strings. 

\section{Complete solution for weight, rope, and shell} \label{Appendix:backreaction}

In the main text I argued that the gravitational backreaction of the supporting rope places a severe constraint on the rate of mining from black holes. In this appendix, I will seek reassurance that there are no further surprises associated with backreaction by deriving the complete solution for the quasistatic mining configuration. The box is modeled as a thin deadweight shell of proper mass $\mathfrak{m}$ at $R_1$. This is supported by a NEC-saturating rope of tension and mass-per-unit-length $\bar{\mu}$. This rope is in turn tethered to a compression structure out at $R_2$ with $S/\mu = w_{\theta}$. Throughout I will assume that the configuration is spherically symmetric. This analysis reduces to that of \cite{Frauendiener:1990} in the limit $\mathfrak{m}, \bar{\mu} \rightarrow 0$. 

The stress-energy of our configuration is given by Eq.~\ref{eq:generalmatter},
\begin{equation}
T^{\mu}_{\ \, \nu} = \textrm{diag} \{- \rho, p_r ,p_{\theta} ,p_{\theta} \} \equiv \frac{1}{4 \pi r^2} \textrm{diag} \{- \mu(r), -T(r) ,S(r),S(r) \}.
\end{equation}
The condition for equilibrium, $\nabla_{\mu} T^{\mu}_{\ \, r} = 0$, is given by Eq.~\ref{eq:underpressure}
\begin{equation}
\frac{dT}{dr} - \frac{1}{\chi} \frac{d \chi}{dr} (\mu - T) + \frac{2}{r} S \label{eq:compressiontoo}
= 0 . 
\end{equation}

Defining $f(r) \equiv ({1 - {2m(r)}/{r} )^{1/2}  } $ then the $G^{t}_{\ \,t} = 8 \pi T^{t}_{\ \,t}$ component of Einstein's Equation is that 
\begin{equation}
\frac{f'}{f} =  \frac{1}{r^2} \frac{m- r \mu}{1 - \frac{2m}{r}} \  \leftrightarrow \ \frac{d m}{dr} = \mu. \label{eq:rrEinstein}
\end{equation}
Combining this with the  $G^{r}_{\ \,r} = 8 \pi T^{r}_{\ \,r}$ component of Einstein's Equation gives 
\begin{equation}
\frac{\chi'}{\chi} =  \frac{1}{r^2} \frac{m- r T}{1 - \frac{2m}{r}}. \label{eq:chiprime}
\end{equation}
Since $f$ and $\chi$ are both normalized to be one at infinity, we see that $f$ and $\chi$ are the same only if surrounded by matter with $T=\mu$, and that $f$ is bigger than $\chi$ for points outside of which there is matter with $\mu > T$. Since the compression structure at $r=R_2$ has $\mu>T$, we will have $f>\chi$ everywhere inside $R_2$.

Now let's consider the metric. 

Inside $R_1$ the metric is
\begin{equation}
ds^2 = - A^2 (1 - \frac{2M}{r} ) dt^2 + \frac{dr^2 }{1 - \frac{2M}{r} } + r^2 d \Omega_2^2.
\end{equation}
Birkhoff's theorem ensures that a black hole surrounded by spherically-distributed matter knows nothing of the matter. We have placed a (locally unobservable) $A$ in there to get the normalization right at infinity: the shell places the hole in a gravitational well relative to infinity and makes time run slow, even slower than results from standard Schwarzschild time dilation. 

Between $R_1$ and $R_2$ the metric is
\begin{equation}
ds^2 = - B^2 (1 - 2 \bar{\mu}) (1 - \frac{2M_2}{r} ) dt^2 + \frac{dr^2 }{(1 - 2 \bar{\mu})(1 - \frac{2M_2}{r} )} + r^2 d \Omega_2^2.
\end{equation}
At $\bar{\mu} = 1/2$ the deficit solid angle reaches $4 \pi$ \cite{Frolov:2001uf} and a horizon forms, as in Eq.~\ref{eq:backreactionofstring}. 

Outside $R_2$ the metric is, again by Birkhoff's theorem,
\begin{equation}
ds^2 = - (1 - \frac{2M_3}{r} ) dt^2 + \frac{dr^2 }{1 - \frac{2M_3}{r} } + r^2 d \Omega_2^2.
\end{equation}

We will use the junction conditions to derive expressions for $M_2$, $M_3$, $A$, $B$  and $\bar{\mu}$ in terms of $\mathfrak{m}$, $M$, $R_1$, $R_2$ and $w_{\theta}$. To keep the expressions short, let us further define two variables $M_a$ and $M_b$ by
\begin{eqnarray}
 (1 - 2 \bar{\mu})(1 -\frac{2M_2}{R_1})  \equiv 1 - \frac{2M_a}{R_1} \  & \leftrightarrow & \ M_a \equiv (1 - 2 \bar{\mu}) M_2 + \bar{\mu} R_1,\\
 (1 - 2 \bar{\mu})(1 -\frac{2M_2}{R_2}) \equiv 1 - \frac{2M_b}{R_2} \ & \leftrightarrow & \ M_b \equiv (1 - 2 \bar{\mu}) M_2 + \bar{\mu} R_2,
\end{eqnarray}
so that $M_a$ is the value of $m(r)$ immediately outside the inner shell, and $M_b$ is the value of $m(r)$ immediately inside the outer shell.

Continuity of the induced metric at $R_1$ gives 
\begin{equation}
\boxed{A^2 (1 - \frac{2M}{R_1}) = B^2 (1 - \frac{2M_a}{R_1})}.
\end{equation}

Continuity of the induced metric at $R_2$ gives 
\begin{equation}
\boxed{B^2 (1 - \frac{2M_b}{R_2}) = (1 - \frac{2 M_3}{R_2} )}.
\end{equation}

At $r=R_1$ the derivative of the metric jumps as it encounters the lump of proper mass $ \mathfrak{m}$. From the definition of the proper mass
\begin{equation}
 d \mathfrak{m} \equiv \mu \, ds = \mu \, dr / f = dm/f, \label{eq:mofmathfrakm}
\end{equation}
we can derive
\begin{equation}
\mathfrak{m} = \int_{M}^{M_a} \frac{dm}{\sqrt{1 - \frac{ 2 m}{R_1}} } = R_1 \left(  \sqrt{1 - \frac{2M}{R_1} } - \sqrt{1 - \frac{2M_a}{R_1} } \right) , \label{eq:mathfrakm}
\end{equation}
which we can rewrite as an expression for $M_a$ as 
\begin{equation}
\boxed{M_a = M + \sqrt{1 - \frac{2M}{R_1} } \mathfrak{m} - \frac{\mathfrak{m}^2}{2R_1} }. \label{eq:thirdjuctioncondition}
\end{equation}

Finally we have the pressure-balance equations across the shells. Using Eq.~\ref{eq:compressiontoo}
\begin{equation}
dT = (\mu - T) \frac{d \chi}{\chi} - \frac{2}{r} S dr ,
\end{equation}
and Eq.~\ref{eq:rrEinstein}
\begin{equation}
dm = \mu dr ,
\end{equation}
and Eq.~\ref{eq:chiprime} 
\begin{equation}
\frac{d \chi}{\chi}  = \frac{1}{r} \frac{m(r) - r T(r)}{r - 2m(r) }   dr,
\end{equation}
and defining $w_\theta \equiv S/ \mu$, which we take to be constant within the shell, gives
\begin{equation}
dT = \left(  \frac{1}{r} \frac{\mu - T}{\mu} \frac{m - r \,T}{r - 2m } - \frac{2}{r} w_\theta  \right) dm.
\end{equation}
Inside a delta-function shell we can approximate $\mu \gg T$ and $r=R$. The solution is
\begin{equation}
T(m) = 1 - \frac{m}{R}+2w_\theta(1 - \frac{2m}{R}) +  c \sqrt{1 - \frac{2m}{R}} .
\end{equation}

At the first junction, at $R=R_1$, the box is assumed to be a deadweight, so $w_\theta=0$. Since $T(m=M)=0$ we have
\begin{equation}
T(m) = 1- \frac{m}{R_1} - \frac{1- \frac{M}{R_1}}{\sqrt{1 - \frac{2M}{R_1}} } \sqrt{1 - \frac{2m}{R_1}} .
\end{equation}
Then since $T(m=M_a) = \bar{\mu}$ we have
\begin{equation}
\boxed{\bar{\mu} = 1- \frac{M_a}{R_1} - \frac{1- \frac{M}{R_1}}{\sqrt{1 - \frac{2M}{R_1}} } \sqrt{1 - \frac{2M_a}{R_1}}} . \label{eq:requiredtensioninstring}
\end{equation}

At the second junction, at $R=R_2$, the rope is tethered to a compression structure with $w_\theta>0$. We can set $c$ by noting that on the outside of the shell we have $T=0$ and $m=M_3$, so  
\begin{equation}
T(m) = 1 - \frac{m}{R_2}+2w_\theta(1 - \frac{2m}{R_2}) - \frac{1 - \frac{M_3}{R_2}+2w_\theta(1 - \frac{2M_3}{R_2})   }{ \sqrt{1 - \frac{2M_3}{R_2}} }  \sqrt{1 - \frac{2m}{R_2}} .
\end{equation}
On the inside of the shell we have $T= \bar{\mu}$ and $m=M_b$, giving our fifth and final constraint
\begin{equation}
\boxed{\bar{\mu}  =  1 - \frac{M_b}{R_2}+2w_\theta(1 - \frac{2M_b}{R_2}) - \frac{1 - \frac{M_3}{R_2}+2w_\theta(1 - \frac{2M_3}{R_2})   }{ \sqrt{1 - \frac{2M_3}{R_2}} }  \sqrt{1 - \frac{2M_b}{R_2}}}. \label{eq:requiredangularstress}
\end{equation}

Let us now explore various limits of these equations. If $\mathfrak{m}=0$ then $\bar{\mu}=0$ and the compression structure has no further weight to bear in addition to its own. In the Schwarzschild limit, $M_3 - M \ll M$, of Eq.~\ref{eq:requiredangularstress} the required angular stress is
\begin{equation}
 w_\theta = \frac{M}{2(r-2M)} \label{eq:requiredw}.
\end{equation}
As promised in Eq.~\ref{eq:wrequired}, a shell at $r=3M$ needs $w_\theta = 1/2$ to support its own weight, and a shell at $ r = 5M/2$ needs $w_\theta=1$.

Finally, we can ask how heavy the box needs to be to cause gravitational collapse. There are in principle two different ways that the box could bring about collapse: it could cause collapse directly by being so heavy that it lies inside the combined Schwarzschild radius of the box plus black hole system; or it could cause collapse indirectly by requiring a supporting rope so heavy that the rope itself collapses. Direct collapse occurs when $M_a=R_1 /2$, which through Eq.~\ref{eq:mathfrakm} corresponds to a proper mass of 
\begin{equation}
\mathfrak{m} =  R_1 \sqrt{1-\frac{2M}{R_1}} .  \label{eq:boxcollapse}
\end{equation}
What about indirect collapse? This happens when the required $\bar{\mu}$ reaches $1/2$, and that happens, Eq.~\ref{eq:requiredtensioninstring} tells us, when $M_a=R_1 /2$, exactly the same criterion as for direct collapse. So for the NEC-saturating ropes envisioned by Eq.~\ref{eq:requiredtensioninstring}, direct and indirect collapse happen simultaneously; for all non-NEC-saturating ropes the weight of the required rope causes collapse before the weight of the box does.

\section{The broadest boxes  saturate the NEC} \label{appendix:stringcaternary}
In Section \ref{sec:TS} we saw that a single NEC-saturating box suspended from a single string near a black hole can be no wider than about a single local Hawking wavelength. In this appendix we will see that this is true not just of all boxes that are NEC-saturating, but also of all boxes that are NEC-satisfying. Indeed we will see that the broadest boxes that obey the NEC also saturate the NEC. 

First let's do the case of a string. Consider a metric
\begin{equation}
ds^2 = - \chi^2 dt^2 + d\chi^2 + d \theta^2 .
\end{equation}
And consider a static matter configuration on this space, so that $T^{\mu}_{\ \nu}(\chi,\theta)$ is not a function of $t$, and $T^{t}_{\ i} = 0$. 
%
The $\nabla_{\mu}T^{\mu}_{\ \chi} = 0$ equation tells us that 
\begin{eqnarray}
\partial_{\chi}T^{\chi}_{\ \chi} + \partial_{\theta}T^{\theta}_{\ \chi}  + \frac{1}{\chi}  \left( T^{\chi}_{\ \chi} -  T^{t}_{\ t} \right) & = & 0 . \label{eq:partialchi}
\end{eqnarray}
The $\nabla_{\mu}T^{\mu}_{\ \theta} = 0$ equation tells us that 
\begin{eqnarray}
\partial_{\chi} T^{\chi}_{\ \theta}  + \partial_{\theta}T^{\theta}_{\ \theta} +  \frac{1}{\chi} T^{\chi}_{\ \theta} & = & 0. \label{eq:partialtheta}
\end{eqnarray}
So far our discussion would apply to a general two-dimensional membrane. But we're interested in codimension-one ropes, which means that we want to be able to cleave this membrane into a sheaf of parallel ropes that do not interact, and then just consider one of these ropes in isolation. This means that the tension perpendicular to the ropes must be zero. Mathematically, this implies that one of the eigenvalues of $T^{\mu}_{\ \nu}$ is zero, so the stress-energy can be written as
\begin{displaymath} 
T^{\mu}_{\ \nu}  = \left( \begin{array}{lll} 
-\rho & 0 & 0 \\ 
0 & p \cos^2 \psi & p\sin \psi \cos \psi \\
0 & p \sin \psi \cos \psi & p \sin^2 \psi 
\end{array} \right), 
\end{displaymath}
where $\psi(\chi,\theta)$ is the angle of the rope to the vertical (to the $\chi$ axis). Vectors \emph{tangent} to the string satisfy
\begin{equation}
\sin \psi \, d \chi - \cos \psi  \, d \theta = 0 , \label{eq:stringtangent}
\end{equation}
whereas vectors \emph{perpendicular} to the string satisfy
\begin{equation}
\cos \psi \, d \chi + \sin \psi \, d \theta = 0 . \label{eq:stringperpendicular}
\end{equation}
And of course as a matter of calculus
\begin{equation}
d\psi =  \partial_\chi \psi \, d \chi  +  \partial_\theta \psi \, d \theta. \label{eq:calculus}
\end{equation}\\
\noindent Taking $\sin \psi$ times Eq.~\ref{eq:partialchi} minus $\cos \psi$ times Eq.~\ref{eq:partialtheta}  gives
\begin{equation}
p (\cos \psi  \ \partial_\chi \psi + \sin \psi \ \partial_{\theta} \psi) = \rho \sin \psi
\end{equation}
or in other words, comparing with Eqs.~\ref{eq:stringtangent} \& \ref{eq:calculus}, 
\begin{equation}
\frac{d \psi}{d \theta} \biggl|_{\textrm{tangent}} = \frac{1}{\chi} \frac{\rho}{p} .
\end{equation}
To simplify comparison with the analysis that led to Eq.~\ref{eq:catenaryform}, we can use $d \psi / d \theta = - \chi''/(1 + {\chi'}^2)$ and $\chi' = \cos \psi/\sin \psi$ to rewrite this as 
\begin{equation}
\frac{\chi''}{1 + \chi'^2} = - \frac{1}{\chi} \frac{\rho}{p} .
\end{equation}
If $\rho=-p$ we get the NEC-saturating shape described in the main text. If $\rho>-p$ the shape is steeper. The NEC forbids $\rho<-p$, so it is impossible for the shape to be any shallower. \\

\noindent For comparison, let's take $\cos \psi$ times Eq.~\ref{eq:partialchi} plus $\sin \psi$ times Eq.~\ref{eq:partialtheta}, giving
\begin{equation}
\sin \psi \, \partial_\theta p  + \cos \psi \,  \partial_\chi p  + p \cos \psi  \, \partial_\theta \psi - p \sin \psi  \, \partial_\chi \psi  = - \frac{1}{\chi}(\rho + p)\cos \psi 
\end{equation}
or in other words, comparing with Eqs.~\ref{eq:stringtangent},~\ref{eq:stringperpendicular} \& \ref{eq:calculus}, 
\begin{equation}
\frac{d p}{d \theta} \biggl|_{\textrm{tangent}}+p \frac{\cos \psi}{\sin \psi} \frac{d \psi}{d \theta} \biggl|_{\textrm{perpendicular}}  =  - \frac{\cos \psi}{\sin \psi} \frac{\rho + p}{\chi} .
\end{equation}
This equation tells us how the tension changes along the string. While the shape of constant-$\mu$ strings has the same functional form as the Newtonian catenary, these equations imply that the coincidence does not extend away from constant-$\mu$ \cite{Fallis}. \\

Now let's do the case of a sheet. Consider a metric
\begin{equation}
ds^2 = - \chi^2 dt^2 + d\chi^2 + d \theta^2 + \theta^2 d \phi^2 .
\end{equation}
And consider a configuration on this space that is both static and rotationally symmetric, so that $T^{\mu}_{\ \nu}(\chi,\theta)$ is not a function of $t$ or $\phi$, and $T^{\chi}_{\ \theta}$ is the only non-zero off-diagonal component. 
The $\nabla_{\mu}T^{\mu}_{\ \chi} = 0$ equation  tells us that 
\begin{eqnarray}
\partial_{\chi}T^{\chi}_{\ \chi}  +  \partial_{\theta}T^{\theta}_{\ \chi} + {  \frac{1}{\theta} T^{\theta}_{\chi} } + \frac{1}{\chi}  \left( T^{\chi}_{\ \chi} -  T^{t}_{\ t} \right)& = & 0 . \label{eq:branepartialchi}
\end{eqnarray}
The $\nabla_{\mu}T^{\mu}_{\ \theta} = 0$ equation tells us that 
\begin{eqnarray}
\partial_{\chi} T^{\chi}_{\ \theta}  + \partial_{\theta}T^{\theta}_{\ \theta} +  \frac{1}{\chi} T^{\chi}_{\ \theta} + {  \frac{1}{\theta} \left( T^{\theta}_{\ \theta} - T^{\phi}_{\ \phi} \right)  }& = & 0. \label{eq:branepartialtheta}
\end{eqnarray}
Note the symmetry between $\chi$ and $\theta$. The stress-energy tensor looks like
\begin{displaymath} 
T^{\mu}_{\ \nu}  = \left( \begin{array}{llll} 
-\rho & 0 & 0 & 0 \\ 
0 & p \cos^2 \psi & p\sin \psi \cos \psi & 0\\
0 & p \sin \psi \cos \psi & p \sin^2 \psi & 0\\
0 & 0 & 0 & q
\end{array} \right), 
\end{displaymath}
where $\psi(\chi,\theta)$ is the angle of the rope to the vertical (to the $\chi$ axis). Vectors \emph{tangent up} to the brane satisfy
\begin{equation}
\sin \psi \, d \chi - \cos \psi  \, d \theta = 0 , \ d \phi = dt = 0,  \label{eq:branetangent}
\end{equation}
 vectors \emph{tangent around} to the brane satisfy
\begin{equation} 
d \theta = d \chi = dt = 0 ,
\end{equation}
whereas vectors \emph{perpendicular} to the brane satisfy
\begin{equation}
\cos \psi \, d \chi + \sin \psi \, d \theta = 0, \ d \phi = dt = 0. \label{eq:braneperpendicular}
\end{equation}
And as a matter of calculus
\begin{equation}
d\psi =  \partial_\chi \psi \, d \chi  +  \partial_\theta \psi \, d \theta. \label{eq:branecalculus}
\end{equation}\\
\noindent Taking $\sin \psi$ times Eq.~\ref{eq:branepartialchi} minus $\cos \psi$ times Eq.~\ref{eq:branepartialtheta}  gives
\begin{equation}
p \left( \cos \psi \, \partial_\chi \psi + \sin \psi \, \partial_\theta \psi \right)  = \frac{\sin \psi}{\chi} \rho + \frac{\cos \psi}{\theta} q,
\end{equation}
or in other words
\begin{equation}
\frac{d \psi}{d \theta} \biggl|_{\textrm{tangent up}} = \frac{1}{\chi} \frac{\rho}{p} + \frac{1}{\theta} \frac{\cos \psi}{\sin \psi} \frac{q}{p} .
\end{equation} 
To simplify comparison with Eq.~\ref{equationofmotion}, we can use $d \psi / d \theta = - \chi''/(1 + {\chi'}^2)$ and $\chi' = \cos \psi/\sin \psi$ to rewrite this as 
\begin{equation}
\frac{\chi''}{1 + \chi'^2} = - \frac{1}{\chi} \frac{\rho}{p}  - \frac{\chi'}{\theta} \frac{q}{p}.
\end{equation}
If $\rho = - p = -q$ we recover the solution of Eq.~\ref{equationofmotion}; any other NEC-obedient choice is narrower. \\

\noindent Finally, let's take $\cos \psi$ times Eq.~\ref{eq:branepartialchi} plus $\sin \psi$ times Eq.~\ref{eq:branepartialtheta}, giving
\begin{equation}
\sin \psi \, \partial_\theta p  + \cos \psi \,  \partial_\chi p  + p \cos \psi  \, \partial_\theta \psi - p \sin \psi  \, \partial_\chi \psi  = - \frac{1}{\chi}(\rho + p)\cos \psi + \frac{ 1}{\theta} q \sin \psi
\end{equation}
or in other words, comparing with Eqs.~\ref{eq:branetangent},~\ref{eq:braneperpendicular} \& \ref{eq:calculus}, 
\begin{equation}
\frac{d p}{d \theta} \biggl|_{\textrm{tangent up}}+p \frac{\cos \psi}{\sin \psi} \frac{d \psi}{d \theta} \biggl|_{\textrm{perpendicular}}  = -  \frac{\cos \psi}{\sin \psi} \frac{\rho + p}{\chi} + \frac{q}{\theta} .
\end{equation}
This equation tells us how the pressure changes along the sheet.

\end{document}